\documentstyle[psfig]{mn}

\title[]{The energy balance of polars revisited}

\author[Ramsay \& Cropper]{
Gavin Ramsay and Mark Cropper\\
Mullard Space Science Laboratory, University College London,
Holmbury St. Mary, Dorking, Surrey, RH5 6NT, UK\\}
\date{Accepted: 18 Sept 2003}

\begin{document}
\outer\def\gtae {$\buildrel {\lower3pt\hbox{$>$}} \over 
{\lower2pt\hbox{$\sim$}} $}
\outer\def\ltae {$\buildrel {\lower3pt\hbox{$<$}} \over 
{\lower2pt\hbox{$\sim$}} $}
\newcommand{\ergscm} {ergs s$^{-1}$ cm$^{-2}$}
\newcommand{\ergss} {ergs s$^{-1}$}
\newcommand{\ergsd} {ergs s$^{-1}$ $d^{2}_{100}$}
\newcommand{\pcmsq} {cm$^{-2}$}
\newcommand{\ros} {\sl ROSAT}
\newcommand{\exo} {\sl EXOSAT}
\newcommand{\xmm} {\sl XMM-Newton}
\def\rchi{{${\chi}_{\nu}^{2}$}}
\def\uchi{{${\chi}^{2}$}}
\newcommand{\Msun} {$M_{\odot}$}
\newcommand{\Mwd} {$M_{wd}$}
\def\Mdot{\hbox{$\dot M$}}
\def\mdot{\hbox{$\dot m$}}
\input psfig.sty

\maketitle

\begin{abstract}

In the {\exo} and {\ros} eras a significant number of polars were
found to show a soft/hard X-ray ratio much greater than that expected
from the standard accretion shock model. This was known as the `soft
X-ray excess'. We have made an snapshot survey of polars using {\xmm}
and determined their soft/hard ratios. We find that less than one in
five of systems show a significant soft X-ray excess, while the rest
show ratios consistent with that predicted by the standard model. We
have investigated the discrepancy between this and the previous
investigations by re-examining all the available {\ros} PSPC pointed
observations of polars using more recent calibrations than in the
original studies.  We find that these data show an energy balance
ratio which is broadly consistent with that of our {\xmm} results. We
conclude that the previous studies were affected by the data being
less well calibrated. We discuss which physical mechanisms might give
rise to a high soft X-ray excess and whether systems with high ratios
show more variation in soft X-rays. Surprisingly, we find that 6 out
of 21 systems found in a high accretion state did not show a distinct
soft X-ray component. Two systems showed one pole with such a
component and one which did not. Based on the ratio of the observed
soft X-ray to UV flux measurements (which were obtained simultaneously
using the Optical Monitor) we suggest that this is because the
reprocessed component in these systems is cool enough to have moved
out of the soft X-ray band and into the EUV or UV band.

\end{abstract}


\section{Introduction}

Polars or AM Her systems are accreting binary systems in which
material transfers from a dwarf secondary star onto a magnetic
($B\sim$10--200MG) white dwarf through Roche lobe overflow. For polars
in a high accretion state, the accretion flow generally forms a strong
shock at some height above the photosphere of the white dwarf. The
maximum temperature in the post-shock flow is set by the mass of the
white dwarf. For a 0.7\Msun white dwarf the shock temperature is
$\sim$30keV, with the temperature decreasing as the gas settles onto
the white dwarf. Some fraction of the hard X-rays intercept the
photosphere of the white dwarf, are thermalised and then re-radiated
as soft X-rays or in the extreme UV. The standard model of the shock
region predicts the ratio of this reprocessed radiation to that
directly emitted by the shock, $L_{reprocessed}/L_{shock}\sim0.5$ (eg
Lamb \& Masters 1979, King \& Lasota 1979). For a recent review of the
physical processes occurring in the shock region see Wu (2000).

Observations made using {\sl EXOSAT} (1983--1986) found that a number
of polars showed a large `soft X-ray excess': if the reprocessed
component was emitted as soft X-rays, the ratio,
$L_{reprocessed}/L_{shock}$, was well in excess of that predicted by
the standard model. In the following decade, Ramsay et al (1994) and
Beuermann \& Burwitz (1995) found using {\sl ROSAT} (1990--1999) data
that many systems showed large excesses. Beuermann \& Burwitz (1995)
calculated the ratio in the {\ros} band (0.1-2.4keV). These authors
suggested that for systems with magnetic field strengths \gtae 30MG,
cyclotron radiation dominates the emission from the shock (Lamb \&
Masters 1979): if this was taken into account then the excess would
largely disappear. On the other hand, Ramsay et al (1994), found that
when bolometric luminosities were used, a significant number of
systems nevertheless showed large soft X-ray excesses.

Various models have been put forward to account for the soft X-ray
excess. These include nuclear burning on the surface of the white
dwarf (Raymond et al 1979, Papaloizou, Pringle \& MacDonald 1982),
accretion energy being transported by electron conduction into the
white dwarf and being re-emitted as soft X-rays (Fabian, Pringle \&
Rees 1976, King \& Lasota 1980, Frank, King \& Lasota 1988) and the
bombardment model (Kuijpers \& Pringle 1982, Thompson \& Cawthorne
1987). None of these models can account for the excess. The most
widely accepted solution to the soft X-ray excess problem is `blobby'
accretion as first proposed by Kuijpers \& Pringle (1982). They
suggested that soft X-rays could be produced by dense blobs of
material which penetrate into the photosphere of the white dwarf, so
that the shock was buried and the hard X-rays emitted by the shock are
thermalised in the photosphere of the white dwarf with the energy
eventually released as soft X-rays. This was developed further by
Frank, King \& Lasota (1988) (see also Litchfield \& King 1990, Frank,
King \& Raine 2002). We also note that soft X-rays are produced near
the base of the post-shock flow (Cropper, Wu \& Ramsay 2000).

However, there is some uncertainty as to which energies the
reprocessed component in the standard model is emitted. Heise \&
Verbunt (1988) found that in the `reversed on-state' the UV and hard
X-ray maxima were sometimes in anti-phase with the soft X-ray minima.
They argued that the reprocessed component is emitted not as soft
X-rays but in the extreme UV. In their scenario, any distinct soft
X-ray component is due to `blobs' of material.

Whilst {\ros} was suited to observing the soft X-ray component, its
lack of sensitivity at higher energies meant that the spectral shape
of the hard X-ray component was not well defined. Further, its energy
resolution was modest. There is now a new generation of X-ray
satellites which combine a high effective area, both at soft and
harder X-ray energies and with higher spectral resolution.

We have undertaken a survey of polars using {\sl XMM-Newton}. This
survey contains observations of 37 polars -- more than half of all
known systems. A preliminary report of the work is given in Ramsay \&
Cropper (2003b) which also provides details of the programmatic
aspects of the survey. A surprising large number of these systems were
found to be in a low accretion state: these observations are described
in Ramsay, Cropper \& Wu (in prep). Detailed investigations into the
phase resolved data of many of these systems have already been
published (see Table 1 for details). This paper reports on the
spectral energy distribution of those systems found to be in a high
accretion state.  We compare these with those results obtained using
{\ros}.

\section{Observations}

{\xmm} was launched in Dec 1999 by the European Space Agency. It has
the largest effective area of any imaging X-ray satellite (Jansen et
al 2001) and also has a 30 cm optical/UV telescope (the Optical
Monitor, OM: Mason et al 2001) allowing simultaneous X-ray and
optical/UV coverage. The EPIC instruments contain imaging detectors
covering the energy range 0.15--10keV with moderate spectra
resolution. Currently the EPIC pn detector (Str\"{u}der et al 2001) is
currently better calibrated at lower energies compared to the EPIC MOS
detector (Turner et al 2001). We therefore restrict the data used in
this paper to the EPIC pn data. The observation log is shown in Table
\ref{log}. By comparison with previous optical and X-ray data we
conclude that all these systems were in high accretion states.

The data were processed using the {\sl XMM-Newton} {\sl Science
Analysis Software} (SAS) v5.3.3. Single and double events were
extracted using an aperture of $\sim40^{''}$ centered on the source
position. Background data were extracted from a source free
region. The background data were scaled and subtracted from the source
data. We used ready-made response files as appropriate for the filter
used (generally `thin').

\begin{table}
\begin{center}
\begin{tabular}{lrcr}
\hline
Source & Rev & Date & Reference \\
\hline
DP Leo & 175 & 2000-11-22 & 1,2\\
WW Hor & 181 & 2000-12-04 & 1,2\\
BY Cam & 314 & 2001-08-26 & 3\\
EU Cnc & 342 & 2001-10-21 & \\
CE Gru & 347 & 2000-10-31 & 4\\
RX J1007--2016 & 365 & 2001-12-07 & 5\\
EV UMa & 366 & 2001-12-08 & 5\\
RX J1002--1925 & 367 & 2001-12-10 & 5\\
V895 Cen & 403 & 2002-02-19 & \\
V347 Pav & 415 & 2002-03-16 & 6 \\
HY Eri & 419 & 2002-03-24 & \\
RX J2115--58& 421 & 2002-03-27 & 7\\
V349 Pav & 438 & 2002-04-30 & \\
AN UMa & 438 & 2002-05-01 & \\
EK UMa & 442 & 2002-05-09 & \\
GG Leo & 444 & 2002-05-13 & 6 \\
EU UMa & 459 & 2002-06-12 & 6 \\
EP Dra & 523 & 2002-09-04 & \\
V1500 Cyg & 531 & 2002-11-02 & \\
VY For & 566 & 2003-01-10 & \\
RX J1846+5538 & 566 & 2003-01-12 & \\
\hline
\end{tabular}
\end{center}
\caption{The sources discussed in the paper: we show the {\xmm}
orbital revolution and date the observations were made. The references
to papers already published using these data are: (1) Ramsay et al
(2001), (2) Pandel et al (2002), (3) Ramsay \& Cropper (2002a), (4)
Ramsay \& Cropper (2002b), (5) Ramsay \& Cropper (2003a), (6) Ramsay
et al (2003), (7) Cropper, Ramsay \& Marsh (2003).}
\label{log}
\end{table}

\section{Spectral Model and other assumptions}
\label{model}

In their analysis of {\ros} data, Ramsay et al (1994) and Beuermann \&
Burwitz (1995) used an emission model consisting of a single
temperature blackbody for the reprocessed component and a single
temperature thermal bremsstrahlung for the post-shock flow (since its
temperature could not be well constrained by the data). The absorption
component was a simple neutral absorption model. These model
components were first approximations to that expected physically.

For this work we improve the model assumptions as follows. The
emission from the hot post-shock region is evidently not a single
temperature: the temperature is hottest near the shock front and the
coolest near the base of the shock. Indeed, a large proportion of the
emission generated in the {\xmm} band will originate in the cooler,
denser region near the photosphere of the white dwarf (Cropper, Wu \&
Ramsay 2000). To model this component we use the multi-temperature
shock model of Cropper et al (1999). In reality the {\it reprocessed}
component arising via irradiation by hard X-rays has a spectral
signature more complex than a simple blackbody (cf Williams, King \&
Brooker 1987, Heise 1995). However, these more physical models are not
available to us. Consequently, we retain the approximation of a
blackbody.  Lastly, the absorption is expected to be more complex than
a neutral absorber (eg Done \& Magdziarz 1998). If neutral absorption
does not to give good fits then an additional absorption component is
added, for instance a partial covering model.

In selecting the data to fit, we have excluded data which correspond
to those orbital phases in which an absorption dip is clearly apparent
(eg in CE Gru, Ramsay \& Cropper 2002b). In addition we have excluded
those phases in which the primary accretion region was not visible.
For DP Leo, WW Hor and BY Cam, we have reprocessed the data using the
more recent version of the SAS than was used by Ramsay et al (2001)
and Ramsay \& Cropper (2002a). The spectra were fitted using {\tt
XSPEC} (Arnaud 1996).

As noted earlier, the standard accretion shock model of Lamb \&
Masters (1979) and King \& Lasota (1979) suggests that
$L_{reprocessed}/L_{shock}\sim0.5$, where $L_{shock}$ is the total
emission from the post-shock region and includes both $L_{X-hard,bol}$
and the cyclotron emission.

Here, we define the hard X-ray luminosity as
($L_{X-hard,bol}=4\pi$Flux$_{X-hard,bol}d^{2}$) where
Flux$_{X-hard,bol}$ is the unabsorbed, bolometric flux from the hard
X-ray component and $d$ is the distance. Since a fraction of this flux
is reflected towards the observer (and our emission model takes into
account the reflection of hard X-rays from the surface of the white
dwarf), we switch the reflected component to zero after the final fit
to determine the intrinsic flux from the optically thin post-shock
region.  We define the soft X-ray luminosity as
($L_{soft,bol}=\pi$Flux$_{soft,bol}$sec($\theta)d^{2}$), where we
assume that the soft X-ray emission is optically thick and can be
approximated by a thin slab of material. The unabsorbed bolometric
flux from the soft X-ray component is Flux$_{soft,bol}$ and $\theta$
is the mean viewing angle to the accretion region.

\section{Results}

\subsection{Systems with no distinct soft X-ray component}

There were several systems whose spectra could be well fitted using an
absorbed multi-temperature shock model, without any requirement for an
additional blackbody component. These are WW Hor, CE Gru, V349 Pav and
V1500 Cyg. Table \ref{fluxhard} gives their observed flux in both the
0.15-10keV and 0.1-2keV bands and also their unabsorbed bolometric
fluxes. 

In two systems (BY Cam and RX J2115--58) accretion at two different
regions was observed, one which showed a soft X-ray component, while
the other did not. It is interesting to note that both these systems
are asynchronous polars (their spin period and orbital periods differ
by a few percent). Ramsay \& Cropper (2002a,b) suggest that a soft
X-ray component is not seen in these systems because their reprocessed
component is cooler compared to the other systems and therefore
shifted to EUV or UV energies. We investigate this further in \S
\ref{nosoft}.

\begin{table}
\begin{center}
\begin{tabular}{lrrr}
\hline
Source & Observed flux & Observed Flux & Flux\\
       & 0.15-10keV & 0.1-2.0keV       & unabs,bol\\ 
       & \ergscm    & \ergscm          & \ergscm  \\
\hline
WW Hor &4.71$\times10^{-13}$ & 1.73$\times10^{-13}$ & 8.80$\times10^{-13}$\\
BY Cam &5.35$\times10^{-12}$ & 2.14$\times10^{-12}$ & 9.03$\times10^{-12}$ \\
CE Gru &3.20$\times10^{-12}$ & 1.02$\times10^{-12}$ &8.23$\times10^{-12}$\\
RX J2115--58 & 7.79$\times10^{-12}$ & 3.05$\times10^{-12}$ &
1.33$\times10^{-11}$\\
V349 Pav & 1.27$\times10^{-12}$ & 3.97$\times10^{-13}$ & 2.64$\times10^{-12}$\\
V1500 Cyg &7.64$\times10^{-14}$ & 2.16$\times10^{-14}$ & 1.41$\times10^{-13}$\\
\hline
\end{tabular}
\end{center}
\caption{The observed fluxes and unabsorbed, bolometric flux for
systems observed using {\xmm} which did not show any evidence for a
distinct soft X-ray component: they could be fitted using a
multi-temperature shock model. In the case of BY Cam and RX J2115--58,
they showed one pole which did show a soft X-ray component and one
which did not.}
\label{fluxhard}
\end{table}

\subsection{Systems with a distinct soft X-ray component}

For those systems which did show both a hard and a soft X-ray
component, we show in Table \ref{flux} the observed flux in both the
0.15--10keV and 0.1-2keV bands (to facilitate comparison with {\sl
ROSAT} observations). We also show the unabsorbed, bolometric fluxes
from the individual model components and the soft-to-hard ratio
(henceforth the `ratio' for brevity) after converting the fluxes to
luminosities (cf \S \ref{model}). We have not taken into account the
geometrical correction to the soft component (which would {\sl
increase} the ratio) nor have we attempted to include an estimate of
the cyclotron component (which would {\sl decrease} the
ratio). Further, we do not take into account reflection of hard X-rays
from the surface of the white dwarf. We also show in Table \ref{flux}
the ratio using a single temperature (30keV) thermal
bremsstrahlung model for the hard component.

\begin{figure}
\begin{center}
\setlength{\unitlength}{1cm}
\begin{picture}(6,20)
\put(-1.1,13.7){\includegraphics{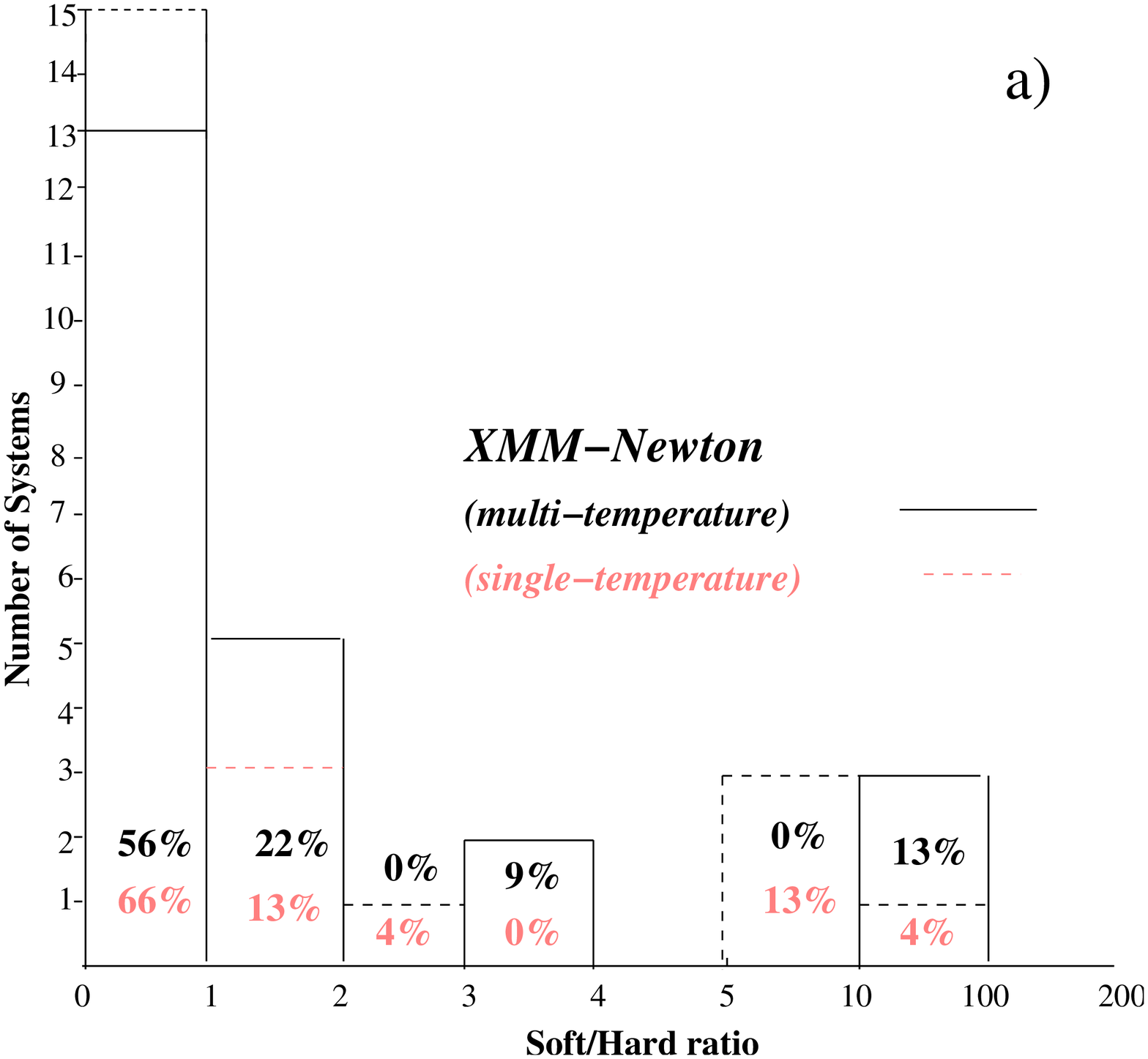}}
\put(-1,10.5){\includegraphics{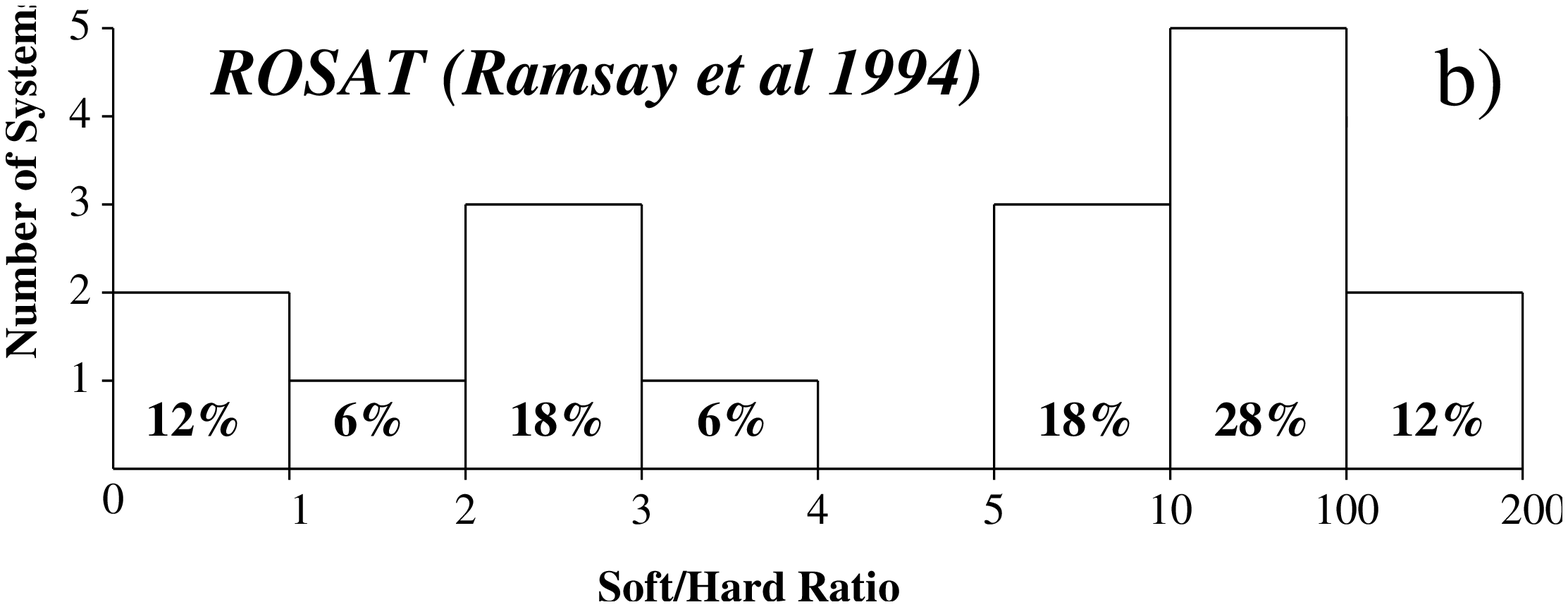}}
\put(-1,6.5){\includegraphics{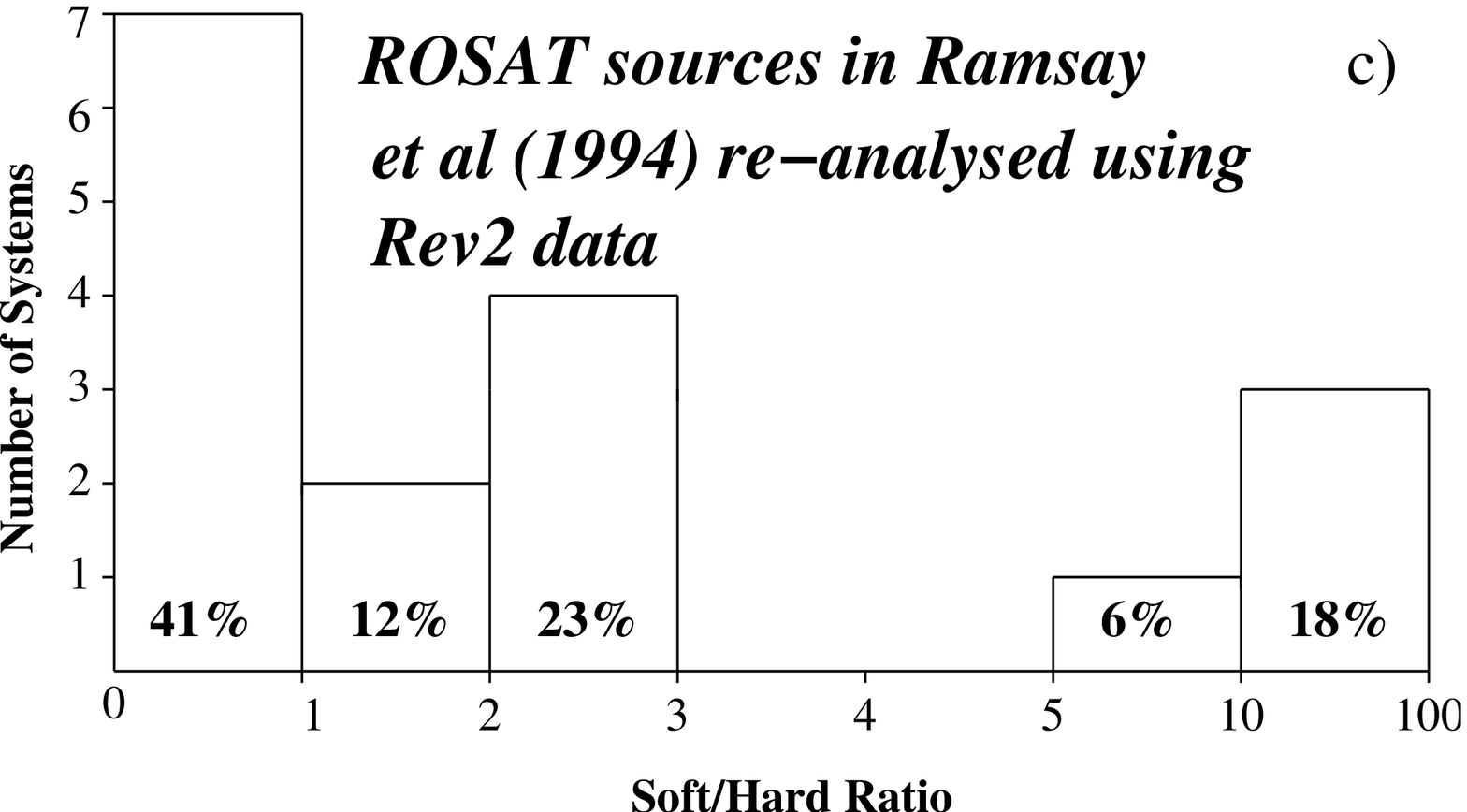}}
\put(-1,0.5){\includegraphics{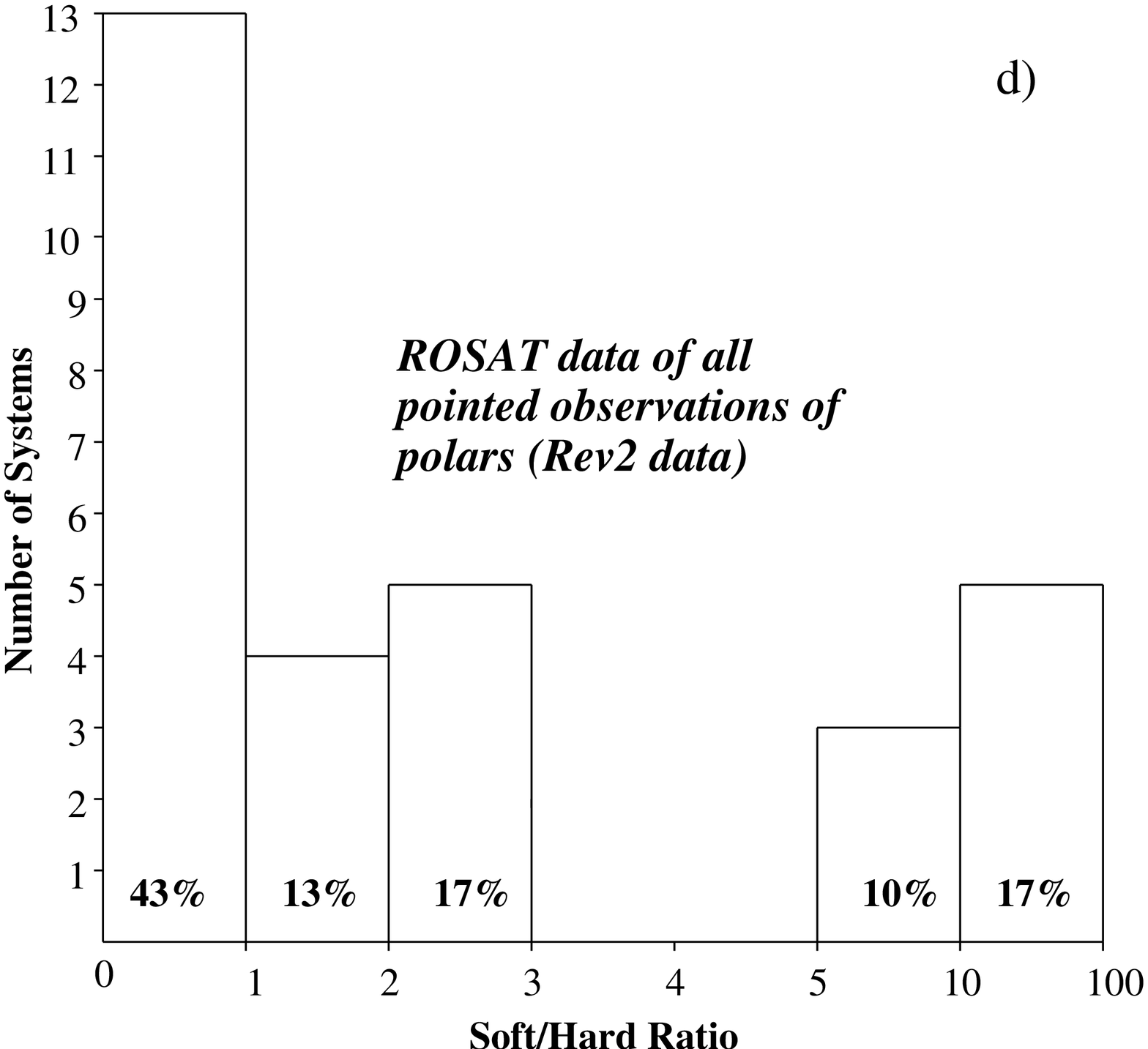}}
\end{picture}
\end{center}
\caption{a): A histogram showing the distribution of the soft/hard
ratio for the polars in a high accretion state in our {\xmm}
survey. We show the results when we use a multi-temperature model for
the shock component and also a single temperature model. We include
those systems (accretion poles) which do not show a distinct soft
X-ray component. b): The distribution for those systems included in
the {\ros} survey of high state systems as determined by Ramsay et al
(1994). c) A histogram showing the distribution of the soft/hard ratio
using {\ros} data for those systems included in the survey of Ramsay
et al (1994) but re-analysed using Rev2 data. d) The energy ratio for
all polars observed using the {\ros} PSPC in a pointed observation
mode.}
\label{ratio}
\end{figure}

The most striking result is the number of systems which show low
ratios. Out of the 17 systems which showed a distinct soft X-ray
component (and modelled using the multi-temperature shock model), 7
have ratios less than 1.0 with another 5 systems having ratios between
1.0 and 2.0. Including those systems in Table \ref{fluxhard} these
numbers increase to 18 out of 23 systems having ratios less than 2.
Once consideration of effects such as reflection of hard X-rays from
the surface of the white dwarf, the correction for optical thickness
effects and the cyclotron component are taken into account, these
systems are likely to be consistent with the standard accretion model.
Only 3 systems show ratios greater than 10 (DP Leo, RX J1007--2016 and
EU UMa). We show this in histogram form in Figure \ref{ratio}a, where
we include those systems which do not show a distinct soft X-ray
component in the smallest ratio bin (therefore BY Cam and RX J2115--58
are `counted twice').

Our survey using {\xmm} data shows that only a small number of polars
show a soft X-ray excess. How does this compare with the findings of
Ramsay et al (1994) who used {\ros} data? Those authors define
$L_{X-hard,bol}=2\pi Flux_{X-hard,bol}d^2$, so we have re-determined
the ratios using our definition of $L_{X-hard,bol}$ described in \S
\ref{model}. Further, Ramsay et al applied a correction to take into
account reflection from the white dwarf; applied a geometrical
correction to the soft X-ray luminosity to account for viewing angle
dependence and made an estimate of the cyclotron luminosity which was
included in $L_{shock}$. To allow ease of comparison with {\xmm} data
and further {\ros} data later in the paper, we make corrections to the
values reported in Ramsay et al (1994) as necessary.  We show a
histogram of the ratio of those systems included in the survey of
Ramsay et al (1994) in Figure \ref{ratio}b. Compared to the results
using {\xmm} these ratios are rather striking: the ratios are more
evenly spread with a bias towards higher ratios.

\begin{table*}
\begin{center}
\begin{tabular}{lrrrrrr}
\hline
Source & Observed Flux & Observed Flux & Soft Flux & Hard
Flux & $L_{s,\pi}/L_{h,4\pi}$ &  $L_{s,\pi}/L_{h,4\pi}$\\
       & 0.15-10keV & 0.1-2keV & unabs,bol & unabs,bol & (MT)& (ST) \\
       & \ergscm & \ergscm & \ergscm & \ergscm & & \\
\hline
DP Leo$^{R}$     & 3.74$\times10^{-13}$ & 4.07$\times10^{-13}$ &
8.42$\times10^{-12}$ & 1.90$\times10^{-13}$ & 11.1 & 8.3\\
BY Cam$^{*R}$ & 2.53$\times10^{-11}$ & 1.10$\times10^{-11}$ &
4.44$\times10^{-11}$ & 4.43$\times10^{-11}$ & 0.25 & 0.1\\
EU Cnc     & 7.90$\times10^{-15}$ & 6.15$\times10^{-15}$ &
6.13$\times10^{-14}$ & 4.60$\times10^{-15}$ & 3.3 & 0.4\\
RX J1007--2016$^{R}$ & 1.70$\times10^{-11}$ & 2.04$\times10^{-11}$ &
2.30$\times10^{-11}$ & 5.74$\times10^{-13}$ & 10.1 & 6.5\\
EV UMa     & 1.65$\times10^{-11}$ & 8.46$\times10^{-12}$ &
3.72$\times10^{-12}$ & 2.63$\times10^{-11}$ & 0.04 & 0.03\\
RX J1002--1925 & 1.81$\times10^{-12}$ & 1.52$\times10^{-12}$ &
1.00$\times10^{-11}$ & 2.35$\times10^{-12}$ & 1.1 & 0.3\\
V895 Cen$^{R}$   & 7.70$\times10^{-13}$ & 6.89$\times10^{-13}$ &
1.49$\times10^{-12}$ & 3.17$\times10^{-13}$ & 1.2 & 1.6\\
V347 Pav$^{R}$   & 1.65$\times10^{-11}$ & 1.05$\times10^{-11}$ &
6.93$\times10^{-11}$ & 2.53$\times10^{-11}$ & 0.7 & 1.1\\
HY Eri$^{R}$ & 1.36$\times10^{-12}$ & 8.42$\times10^{-13}$ &
2.67$\times10^{-11}$ & 2.02$\times10^{-12}$ & 3.3 & 2.3\\ 
RX J2115--58$^{*R}$ & 1.89$\times10^{-11}$ & 8.14$\times10^{-12}$ &
8.02$\times10^{-12}$ & 3.93$\times10^{-11}$ & 0.05 & 0.1\\ 
AN UMa$^{R}$     & 6.38$\times10^{-12}$ & 6.12$\times10^{-12}$ &
1.34$\times10^{-11}$ & 6.07$\times10^{-12}$ & 0.6 & 0.7\\
EK UMa$^{R}$     & 3.91$\times10^{-12}$ & 3.58$\times10^{-12}$ &
2.86$\times10^{-11}$ & 5.20$\times10^{-12}$ & 1.2 & 6.1\\
GG Leo$^{R}$     & 1.15$\times10^{-11}$ & 4.14$\times10^{-12}$ &
2.09$\times10^{-11}$ & 3.03$\times10^{-11}$ & 0.2 & 0.1\\
EU UMa$^{R}$     & 6.86$\times10^{-12}$ & 1.37$\times10^{-11}$ &
2.33$\times10^{-10}$ & 5.09$\times10^{-13}$ & 116 & 33\\
EP Dra     & 5.45$\times10^{-12}$ & 2.15$\times10^{-12}$ &
7.13$\times10^{-12}$ & 1.30$\times10^{-11}$ & 0.1 & 0.4\\
VY For     & 2.74$\times10^{-13}$ & 2.53$\times10^{-13}$ &
1.18$\times10^{-12}$ & 1.62$\times10^{-13}$ & 1.8 & 0.9\\ 
RX J1846+55$^{R}$ & 2.09$\times10^{-12}$ & 1.05$\times10^{-12}$ &
2.77$\times10^{-11}$ & 3.78$\times10^{-12}$ & 1.8 & 1.5\\
\hline
\end{tabular}
\end{center}
\caption{The observed fluxes in the 0.15-10keV and 0.1-2.0keV bands
using an absorbed blackbody plus multi-temperature model (MT) using
{\xmm} data. We also show the unabsorbed, bolometric fluxes (unabs,
bol) derived for the soft X-ray and the hard X-ray components using a
MT model. Those systems marked with a $^*$ were observed to have two
accretion poles: one which showed a soft component and one which did
not. These fluxes refer to the `soft' pole. We show in the last two
columns, the ratio of the luminosities derived using the MT model and
also a model where we assume a single temperature (30keV) thermal
bremsstrahlung model (ST) for the shock component. We have not taken
into account the geometrical correction in determining the soft X-ray
luminosity nor the cyclotron component or reflection of hard X-rays
from the surface of the white dwarf. $^{R}$ refers to those systems
which were observed using {\ros} in the PSPC pointed programme and in
a high accretion state.}
\label{flux}
\end{table*}

\section{Comparing the {\xmm} results with those of {\ros}}

There are several possible reasons for the discrepancy.  We examine
the affect of the model which is used to fit the data and the
intensity of the source. In addition, since the studies of Ramsay et al
(1994) and Beuermann \& Burwitz (1995) the calibration of {\ros} data
is more mature: we revisit the available {\ros} data on polars to
determine if we confirm the conclusions of these earlier studies.

\subsection{The effect of the model used on the ratio}

As noted before, in the {\ros} studies, the spectra were fitted using
a simple absorbed blackbody plus single temperature thermal
bremsstrahlung emission model. We have therefore fitted the {\xmm}
spectra using this model. We fix the temperature of the hard X-ray
component at 30keV as was typically used in the {\ros} studies and add
a Gaussian line near 6.7keV to account for any Fe K$\alpha$
emission. We show in the last column of Table \ref{flux} the energy
ratio using this model. We show the results in histogram form in
Figure \ref{ratio}a which also shows the ratio using the
multi-temperature model for the shock component.

Comparing the ratio using this single temperature shock model with
that found using the multi-temperature shock model, we find that many
systems show similar ratios. There appears to be no consistent
tendency for one model to give ratios which are consistency larger or
smaller than the other. Those systems giving larger differences, EU
Cnc, EK UMa and EU UMa, can be accounted for, in the case of the first
system, its comparative faintness, in the second the fact that a
single temperature model gave a rather poor fit, and in the third
system the ratio is very large so that even a small difference in the
fitted absorption will affect the resulting unabsorbed, bolometric
flux. We conclude that the model which is used to fit the X-ray data
has, generally, a small effect on the resulting ratio.

Ramsay \& Cropper (2003b) investigated the affect of adding a more
complex absorption model (neutral absorption plus partial covering
model) and found that this had a greater affect on the derived ratio
compared with varying the emission model.

\subsection{The effect  of the accretion state on the ratio}
\label{accretionstate}

\begin{figure}
\begin{center}
\setlength{\unitlength}{1cm}
\begin{picture}(6,5.5)
\put(-0.5,-0.8){\includegraphics{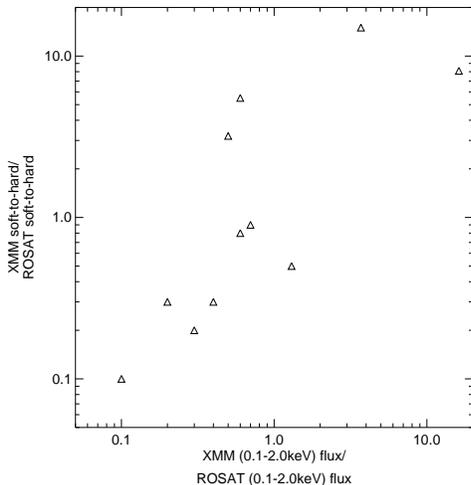}}
\end{picture}
\end{center}
\caption{The ratio of the observed 0.1--2.0keV flux determined using
{\xmm} and {\ros} against the energy balance ratio determined using
{\xmm} and {\ros} for a given source.}
\label{flux_ratio}
\end{figure}

We know that the ratio is dependent on the brightness of the system:
using {\ros} observations of AM Her, the ratio was found to be
significantly higher when it was brighter and BL Hyi showed higher
ratios during flares (Ramsay, Cropper \& Mason 1995).  To further
explore the effects of brightness level on the ratio, we extracted
data from the {\ros} archive for our {\xmm} sources (those shown in
Table \ref{flux}).  We extracted data in a similar way to that of our
{\xmm} data: namely, restricting the data to cover phase intervals
covering the bright phase and excluding phases of absorption dips or
eclipses. For simplicity, and make it easy for others to reproduce, we
fit the spectra using an absorbed blackbody plus thermal
bremsstrahlung of fixed temperature (30keV). We derived the observed
flux in the 0.1--2.0keV band from both the {\xmm} and {\ros}
observations. We also determined the unabsorbed, bolometric
luminosities of both the soft and hard X-ray components using these
data and compute their ratio. We show the relationship between the
brightness and the energy balance ratio in Table \ref{xmm_rosat_flux}
and Figure \ref{flux_ratio}. We find that when a given system is
bright, the trend is for higher energy balance ratios. This is
consistent with that found using the {\ros} data.

All subsets of a population can be biased. For instance, if one sample
was biased towards less luminous polars then the results shown in
Figure \ref{flux_ratio} imply that this sample would be biased towards
lower ratios. In addition, a number of bright and well observed polars
were not included in our {\xmm} survey - eg AM Her, VV Pup, UZ For.

To investigate this further, we determined the bolometric X-ray
luminosity of each system included in the {\xmm} and {\ros}
surveys. We used the best available distance estimate in the
literature (for those systems where no distance estimate was available
we assumed a canonical distance of 100pc). For brevity, we do not
record the luminosities and distances for each system. Suffice to say,
the mean bolometric X-ray luminosity for our {\xmm} survey is
2$\times10^{32}$ \ergss and for the {\ros} sample 3$\times10^{32}$
\ergss. The mean luminosity of the {\ros} sample of Ramsay et al
(1994) was 2$\times10^{32}$ \ergss. This implies that the overall
samples were not influenced in such a way that the intrinsic
luminosity biased the resulting distribution of ratios (of course,
this is not the case for any individual system).

We also compared the total luminosity of a system with its energy
balance ratio. We find that there is no clear relationship between
these quantities. This suggests that while an increase in the
accretion rate will give a higher ratio for any given polar (assuming
the accretion rate is proportional to the total X-ray luminosity), a
given accretion rate will not produce a certain ratio in every
system. There are clearly additional factors which influence the ratio.

\begin{table}
\begin{center}
\begin{tabular}{lrr}
\hline
Source & {\sl XMM}/ & {\sl XMM}/ \\ 
       & {\ros} &   {\ros} \\
       & Flux & ($L_{s,\pi}/L_{h,4\pi}$)\\
\hline
DP Leo &  0.2 &  0.3 \\
BY Cam &  1.3 &  0.5 \\
RX J1007--2016&  16.2  & 8.1 \\
V895 Cen &  0.5 &  3.2 \\
V347 Pav &  0.6 &  5.5 \\
HY Eri & 0.1 & 0.1 \\
AN UMa &  0.7 &  0.9 \\
EK UMa &  0.4 &  0.3 \\
GG Leo &  0.3 &  0.2 \\
EU UMa &  3.7 &  15.0 \\
RX J1846+5537 & 0.6 & 0.8 \\
\hline
\end{tabular}
\end{center}
\caption{Those sources in our {\xmm} survey which were also observed
in the {\ros} PSPC pointed programme. We show the ratio of the flux
observed using {\xmm} compared to that found using {\ros}.  We also
show the ratio of the energy balance ratio determined using {\xmm} and
{\ros} (assuming a single temperature (30keV) component for the
post-shock region).}
\label{xmm_rosat_flux}
\end{table}

\subsection{The effect of the calibration on the ratio}

The original studies of {\ros} data by Ramsay et al (1994), Beuermann
\& Burwitz (1995) and others, were made using `Rev0' calibrated
data. Further, the response files which were used were less definitive
than currently available. {\ros} data have now been reprocessed
several times, with `Rev2' data now available in the public
archives. In addition, for soft X-ray sources (thus including polars)
a higher detector response was found compared to the original response
files, while an excess was found at higher energies (cf calibration
documents in the MPE web site).  We have therefore retrieved all
available {\ros} data of polars in the public archive. We extracted
data in a similar manner to that described in \S \ref{accretionstate}.

Firstly, we consider those 17 sources reported by Ramsay et al (1994)
which were found to be in a high accretion state.  Of those sources
(7) in which upper limits were placed on the bremsstrahlung X-ray flux
in Ramsay et al (1994), we now find, using Rev2 data, that the
spectral fits are significantly worse if we do not include a thermal
bremsstrahlung. We then compared the observed fluxes in the 0.1-0.4keV
and 0.5-2.0keV bands derived using the blackbody and thermal
bremsstrahlung components respectively and compared them directly with
those fluxes reported by Ramsay et al (1994) (the spectral files used
by Ramsay et al 1994 are not available). We then compared the ratio
0.1-0.4keV/0.5-2.0keV determined using Rev0 and Rev2 data (henceforth
Rev0/Rev2). 

There were 8 sources which were positioned off-axis: of these 5 had
upper limits to the thermal bremsstrahlung flux over 0.5--2.0keV in
the Rev0 analysis, while this component was required for a good fit
using Rev2 data. The remaining three had a mean ratio
Rev0/Rev2=2.0. Therefore, for off-axis sources, there is some
indication that the soft X-ray flux was more prominent in the Rev0
data than in the Rev2 data. For the 9 on-axis source 2 had upper
limits to the thermal bremsstrahlung flux in the 0.5-2.0keV band in
the Rev0 data. For the remaining 7 sources, the situation is less
clear with some systems having higher Rev0/Rev2 ratio while it was
lower in others - the mean was 1.1.  These results suggest that the
energy balance determined using {\ros} data should be
re-examined. (Ramsay \& Cropper 2003b focused on {\ros} and {\xmm}
observations of AN UMa and found a much higher ratio using {\ros} data
compared with {\xmm} data. For the {\ros} analysis we used an
inappropriate response file so that particular result in that paper
should be disregarded).

We determined the observed flux in the 0.1--2.0keV band and the
bolometric fluxes of the soft and hard X-ray components. We also
compute the soft to hard luminosity ratio (cf \S \ref{model}). The
results are shown in Table \ref{rosat} and in histogram form in Figure
\ref{ratio}c. It is clear that using the Rev2 data, the energy balance
ratio is now biased towards lower ratios compared with those
determined using Rev0 data and more like the {\xmm} distribution.

We now include all those polars which were observed in a pointed PSPC
observation and not included in the survey of Ramsay et al (1994):
these results are also shown in Table \ref{rosat} and in Figure
\ref{ratio}d. Taken as a whole, the {\ros} observations of polars show
a similar result to that found using our {\xmm} data: the ratios are
biased towards low values, with only $\sim$1/4 of systems showing high
ratios (slightly more than found with our {\xmm} data (Figure
\ref{ratio}a)). We conclude that the reason why Ramsay et al (1994)
found many systems to have a high ratios is due to the data being less
well calibrated than is now possible.

\begin{table*}
\begin{center}
\begin{tabular}{lrrrrr}
\hline
Source  & Start & Observed   & Soft flux & Hard Flux & $L_{s,\pi}/L_{h,4\pi}$\\
        & Date  & 0.1-2.0keV & \ergscm   & \ergscm   & \\
\hline
BY Cam$^{X}$ & 1991-03-09 & $8.43\times10^{-12}$ & $3.47\times10^{-11}$ & 
$3.52\times10^{-11}$& 0.2 \\
V834 Cen$^{XL}$ & 1992-07-27 & $2.76\times10^{-11}$ & $6.98\times10^{-11}$ & 
$2.42\times10^{-11}$& 0.7 \\
EF Eri & 1990-07-18 & $4.16\times10^{-11}$ & $3.45\times10^{-11}$ & 
$1.84\times10^{-10}$& 0.05 \\
UZ For$^{XL}$ & 1991-08-14 & $1.67\times10^{-11}$ & $2.84\times10^{-11}$ & 
$2.96\times10^{-12}$& 2.4 \\
AM Her & 1991-04-12 & $3.37\times10^{-10}$ & $4.49\times10^{-10}$ & 
$1.16\times10^{-10}$& 1.0 \\
BL Hyi & 1991-04-15 & $2.42\times10^{-11}$ & $1.89\times10^{-10}$ & 
$1.66\times10^{-11}$ & 2.9 \\
DP Leo$^{X}$ & 1993-05-30 & $1.96\times10^{-12}$ & $6.48\times10^{-12}$ & 
$6.47\times10^{-14}$& 25 \\
VV Pup & 1991-10-17 & $1.28\times10^{-10}$ & $1.90\times10^{-10}$ & 
$1.76\times10^{-11}$& 2.6 \\
AN UMa$^{X}$ & 1991-12-04 & $1.04\times10^{-11}$ & $2.24\times10^{-11}$ & 
$6.91\times10^{-12}$& 0.8 \\
EK UMa$^{X}$ & 1992-05-12 & $6.31\times10^{-12}$ & $9.12\times10^{-12}$ & 
$1.10\times10^{-13}$& 21 \\
QQ Vul & 1991-04-12 & $6.28\times10^{-12}$ & $8.70\times10^{-12}$ & 
$6.35\times10^{-12}$& 0.3 \\
RS Cae$^{XL}$ & 1992-09-21 & $3.50\times10^{-12}$ & $1.46\times10^{-11}$ & 
$6.55\times10^{-13}$& 5.5 \\
RX J1007-2016$^{X}$ & 1992-11-17 & $1.58\times10^{-12}$ & $3.24\times10^{-12}$ & 
$1.02\times10^{-12}$& 0.8 \\
EU UMa$^{X}$ & 1993-05-25 & $3.66\times10^{-12}$ & $9.14\times10^{-12}$ & 
$1.06\times10^{-12}$& 2.2 \\
V347 Pav$^{X}$ & 1993-03-26 & $1.81\times10^{-11}$ & $3.77\times10^{-11}$ & 
$4.81\times10^{-11}$& 0.2 \\
QS Tel & 1992-10-13 & $5.17\times10^{-12}$ & $1.32\times10^{-11}$ & 
$2.68\times10^{-12}$& 1.2 \\
HU Aqr & 1993-10-27 & $4.09\times10^{-11}$ & $3.14\times10^{-10}$ & 
$1.04\times10^{-10}$& 11 \\
\hline
CV Hyi$^{XL}$ & 1993-11-03 & $1.30\times10^{-13}$ & $5.86\times10^{-13}$ & 
$1.37\times10^{-13}$& 1.1 \\
RX J0153--59$^{XL}$& 1992-07-01& $5.08\times10^{-13}$ & $1.12\times10^{-12}$ & 
$1.92\times10^{-12}$& 0.2 \\
AI Tri & 1993-07-19& $1.79\times10^{-12}$ & $8.58\times10^{-12}$ & 
$3.05\times10^{-12}$& 0.7 \\
HY Eri$^{XL}$ & 1992-02-24& $6.68\times10^{-12}$ & $9.35\times10^{-10}$ & 
$<4.34\times10^{-13}$& $>$58 \\
RX J0803-47 & 1991-12-03& $1.59\times10^{-12}$ & $3.20\times10^{-11}$ & 
$<1.54\times10^{-13}$& $>$53 \\
MN Hya$^{XL}$ & 1994-06-01& $3.80\times10^{-13}$ & $6.87\times10^{-12}$ & 
$5.58\times10^{-12}$& 0.3 \\
V381 Vel & 1993-11-30& $9.14\times10^{-13}$ & $5.51\times10^{-12}$ & 
$2.28\times10^{-12}$& 6.0 \\
FH UMa$^{XL}$ & 1993-05-12& $3.15\times10^{-13}$ & $1.56\times10^{-12}$ & 
$<7.36\times10^{-14}$& $>$5.3 \\
V884 Her & 1993-09-11& $1.19\times10^{-11}$ & $6.30\times10^{-11}$ & 
$7.56\times10^{-12}$& 2.1\\
V895 Cen$^{XL}$  & 1997-02-12& $1.30\times10^{-12}$ & $5.09\times10^{-12}$ & 
$3.23\times10^{-12}$& 0.5\\
GG Leo$^{X}$ & 1993-11-09& $1.47\times10^{-11}$ & $4.33\times10^{-11}$ & 
$2.11\times10^{-11}$& 0.5\\
RX J1846+5537$^{X}$ & 1992-06-15& $1.76\times10^{-12}$ & 
$2.40\times10^{-11}$ & $3.14\times10^{-12}$& 1.9\\
V1432 Aql & 1993-03-31 & $9.08\times10^{-12}$ & $7.86\times10^{-11}$ & 
$4.66\times10^{-11}$& 0.4 \\
\hline
\end{tabular}
\end{center}
\caption{The log of all pointed {\ros} PSPC observations of
polars. The date shown is the start date of the observations. We show
the observed flux in the 0.1-2.0keV band as well as the bolometric
luminosity derived from the soft X-ray and hard X-ray components. We
model the spectra using an absorbed blackbody plus thermal
bremsstrahlung component ($kT$=30keV). The top panel gives those
polars contained in the work of Ramsay et al (1994), while the bottom
panel were not. The $^{X}$ refers to those systems which were part of
our {\xmm} survey and in a bright state (Table \ref{flux}), while
$^{XL}$ refers to those systems which were part of our {\xmm} survey
and in a low accretion state (Ramsay, Cropper \& Wu, in prep).}
\label{rosat}
\end{table*}

\section{The status of the soft X-ray excess in polars}

There is a general impression that there is a prevalent soft X-ray
excess problem in polars. In his book `{\sl Cataclysmic Variable
stars}' (Warner 1995), Warner states `...lead to the conclusion that a
soft X-ray excess does exist.'. Similarly, in their book `{\sl
Accretion Power in Astrophysics}', Frank, King \& Raine (2002) claim
that `the $L_{[hard]}/L_{soft}$ is always found to be much smaller
than [predicted]'.

We conclude from our {\xmm} observations and from our re-analysis of
{\ros} PSPC data that most (but not all) systems show a low soft/hard
ratio.  Once consideration of affects such as reflection of hard
X-rays from the surface of the white dwarf, the correction for optical
thickness effects and the cyclotron component are taken into account
(we do not do this here as for many systems these corrections are
unknown), most systems are likely to be consistent with the standard
accretion model. A relatively small number of systems do, however,
show a strong excess.  This is in contrast to the earlier studies
which showed that many systems had medium to high ratios.

As noted in our introduction, the most likely cause of the soft X-ray
excess is blobs of material. This was called into question by Greeley
et al (1999) who observed AM Her using the Hopkins UV telescope and
found evidence for rapid variations in flux. They suggested their
observations were not consistent with the blobby accretion model: they
determined the average density of the flares assuming that the mass in
the flare was deposited over the extent of the UV hot-spot. This was
much less than the minimum density required for a `blob' of material
to form a buried shock below the photosphere of the white
dwarf. However, King (2000) argued that the data are consistent since
Greeley et al (1999) assumed that the area over which accretion occurs
($A_{acc}$), and the total area over which the reprocessed emission
emerges ($A_{eff}$) are the same, while in actual fact,
$A_{acc}<<A_{eff}$. Therefore, King (2000) concludes that blobby
accretion is still the most viable solution to account for the soft
X-ray excess.  We now go on to determine whether there is any common
factor for those, relatively few, systems which show a high ratio.

\section{Why a high ratio?}

\subsection{Orbital Period}

There are 3 systems which show high ratios using {\xmm} data: DP Leo,
RX J1007-2016 and EU UMa. Further, using {\ros} data there were 5
systems showing high ratios: DP Leo, EK UMa, HY Eri, RX J0803--47 and
HU Aqr. It is clear that the orbital period is not a factor in
determining the ratio: DP Leo and EU UMa have orbital periods of
$\sim$90 mins while RX J1007--2016 has a period of 208 mins.

\subsection{Magnetic Field}

\begin{figure}
\begin{center}
\setlength{\unitlength}{1cm}
\begin{picture}(8,9.5)
\put(-0.5,-0.8){\includegraphics{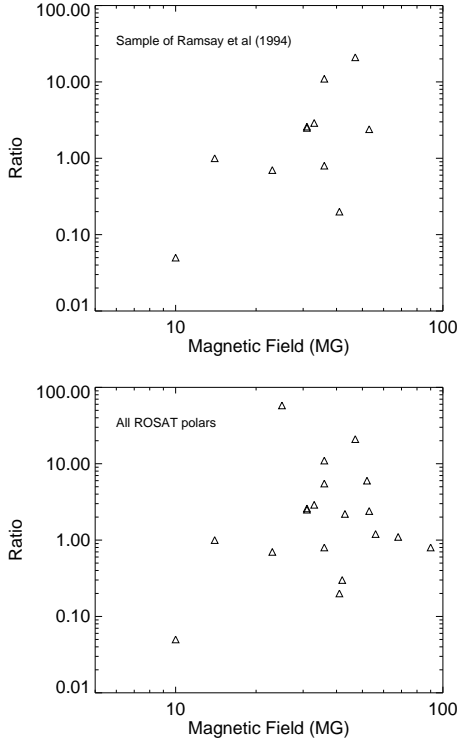}}
\end{picture}
\end{center}
\caption{The energy balance ratio as a function of magnetic field
strength for those systems in the study of Ramsay et al (1994) (left),
and all those polars observed in the pointed mode (Rev2 data was
used).}
\label{field}
\end{figure}

The second parameter which we consider is the magnetic field
strength. RX J1007-2016 with a high ratio has a magnetic field of
$\sim$90 MG (Reinsch et al 1999), while DP Leo has a magnetic field
strength of 31MG for its primary (X-ray emitting) accretion pole
(Cropper \& Wickramasinghe 1993). There is no estimate for the field
strength of EU UMa in the literature. However, systems which show
similarly high magnetic field strengths also show low ratios: for
instance, BY Cam (41MG), AN UMa (36MG) and EK UMa (47MG). This is in
contrast to the results of Ramsay et al (1994) who found a trend
between the ratio and the magnetic field, with systems with low
magnetic fields tending to have low ratios. We show in Figure
\ref{field} the ratio derived using Rev2 {\ros} data against magnetic
field strength for those systems in the sample of Ramsay et al (1994)
which had magnetic field estimates: this shows a rather weak trend.
We also show these values using all systems observed using {\ros}: the
complete {\ros} sample shows no evidence of a trend. We conclude that
the trend of energy balance ratio against magnetic field was largely
due to the small sample which was used.

\subsection{The coupling radius and stream field orientation}

A third parameter which may affect the ratio is the point at which the
accretion flow feels the force of the magnetic field of the white
dwarf: the coupling radius, $R_{c}$.  Although the location of this
point is difficult to determine theoretically, we use the same
relationship as was used by Ramsay et al (1994): 

$R_{c}= r_{\mu}\cos^{2}(\beta) = \cos^{2}(\beta) 5.1\times10^{8}
M_{1}^{-1/7} \dot{M}_{16}^{-2/7} \mu_{30}^{4/7}$ cm

where $\beta$ is the angle between the spin and magnetic axis and the
other parameters their usual meanings.  Since the mass transfer rate
is generally poorly known (being dependent on distance) we take the
observed orbital period-mass transfer rate to set this parameter
(Warner 1995). For magnetic field strengths we use the value in
Wickramasinghe \& Ferrario (2000), or when not known we assume
$B$=20MG. For the mass of the white dwarf we take the values from the
literature or assume $M_{wd}$=0.7\Msun when not known. We find that
the energy balance ratio determined using {\xmm} data is not
correlated with $R_{c}$. This is in contrast to Ramsay et al (1994)
who found using the Rev0 {\ros} data that low ratio systems tended to
have small coupling radii, and high ratio systems tended to have large
coupling radii. We conclude this trend was probably due to chance,
although we discuss below the reasons why $\beta$ may play some role
in determining the resulting X-ray spectrum.

The angle $\beta$ may also affect the characteristics of the accretion
flow and therefore the X-ray spectrum. If $\beta$ is at high angles,
the accretion flow travels further before feeling the effect of the
magnetic field. This may influence the resulting X-ray
spectrum. Unfortunately, this value is not well known for many
systems. However, DP Leo has $\beta=103^{\circ}$, EK UMa
$\beta=56^{\circ}$ (Cropper 1990) and both show high ratios. In
contrast, HU Aqr, has a relatively small dipole offset
($\beta=16^{\circ}$, Schwope et al 2001), but shows a high ratio
(although smaller than DP Leo and EK UMa). Clearly, polarimetric
observations of the other systems showing high ratios (HY Eri and RX
J0803--47) are encouraged so that their accretion geometry can be
determined.

We now consider the angle that the accretion stream makes with the
magnetic axis at the point where it couples onto the magnetic field of
the white dwarf. The azimuth of the magnetic axis tends to point ahead
of the line center joining the binary components, with a mean angle of
$\sim20^{\circ}$ (Cropper 1988). In contrast, the coupling point is
known to vary from cycle to cycle and is related to the accretion rate
(eg Bridge et al 2002, 2003). Small changes in the coupling point
could make large changes to the angle that the stream makes with the
magnetic axis. Could this affect the resulting X-ray spectrum? The
asynchronous polars have a changing stream-magnetic field orientation,
repeating once every beat period. We observed two of them -- RX
J2115--58 and BY Cam. We find that they both show one pole which shows
a distinct soft X-ray component and one pole which does not. This
implies that the stream-field orientation does have an affect on the
resulting X-ray spectrum. We have {\xmm} observations of RX J2115--58
which were made at 7 different stream-field orientations. A brief
summary of these observations are reported by Cropper, Ramsay \& Marsh
(2003). Detailed work is in progress to determine how this orientation
affects the resulting X-ray spectra and the resulting energy balance
ratio.

\section{Short time variability}

We now investigate if the degree of variability in soft X-rays is
related to the energy balance ratio. Simple considerations suggest
that systems with high ratios may be expected to show enhanced soft
X-ray variability since the most likely physical mechanism for
producing the soft X-ray excess is that of dense blobs of material
(Kuijpers \& Pringle 1982).

We extracted events in the energy range 0.15--0.5keV (since the soft
component lies in this band) from a narrow radius around the source --
again excluding faint phases, absorption dips or eclipses.  There are
various techniques to measure the variability of an object: we now
discuss several of these.

Although a weak test of source flickering behaviour, we first
performed a Discrete Fourier Transform of these events. There was no
correlation between the power spectrum (for instance the frequency at
which there was an upturn in power) and the energy ratio which we
measured earlier. Secondly, we determined the time between events and
divided this time by the mean time separation of events (to account
for the brightness of the source) and made a histogram of this time
difference. Again there was no correlation between the shape of the
histograms and the energy ratio.

We then binned the event rate files into 10 sec bins and then measured
the mean and the standard deviation of these light curves. In those
systems where there was a trend in the light curve, we remove this
trend using a polynomial fit. We find no correlation between the
standard deviation and the energy balance ratio. We also compared the
expected variation in the light curve, $\sigma_{e}$, with that
measured, $\sigma_{o}$: the more variable the source the higher the
ratio $\sigma_{o}/\sigma_{e}$ (Sokoloski, Bildstein \& Ho 2001). This
was done for various bin sizes: we find no trend with the energy
balance ratio.

\begin{figure}
\begin{center}
\setlength{\unitlength}{1cm}
\begin{picture}(8,6.5)
\put(-0.7,0){\includegraphics{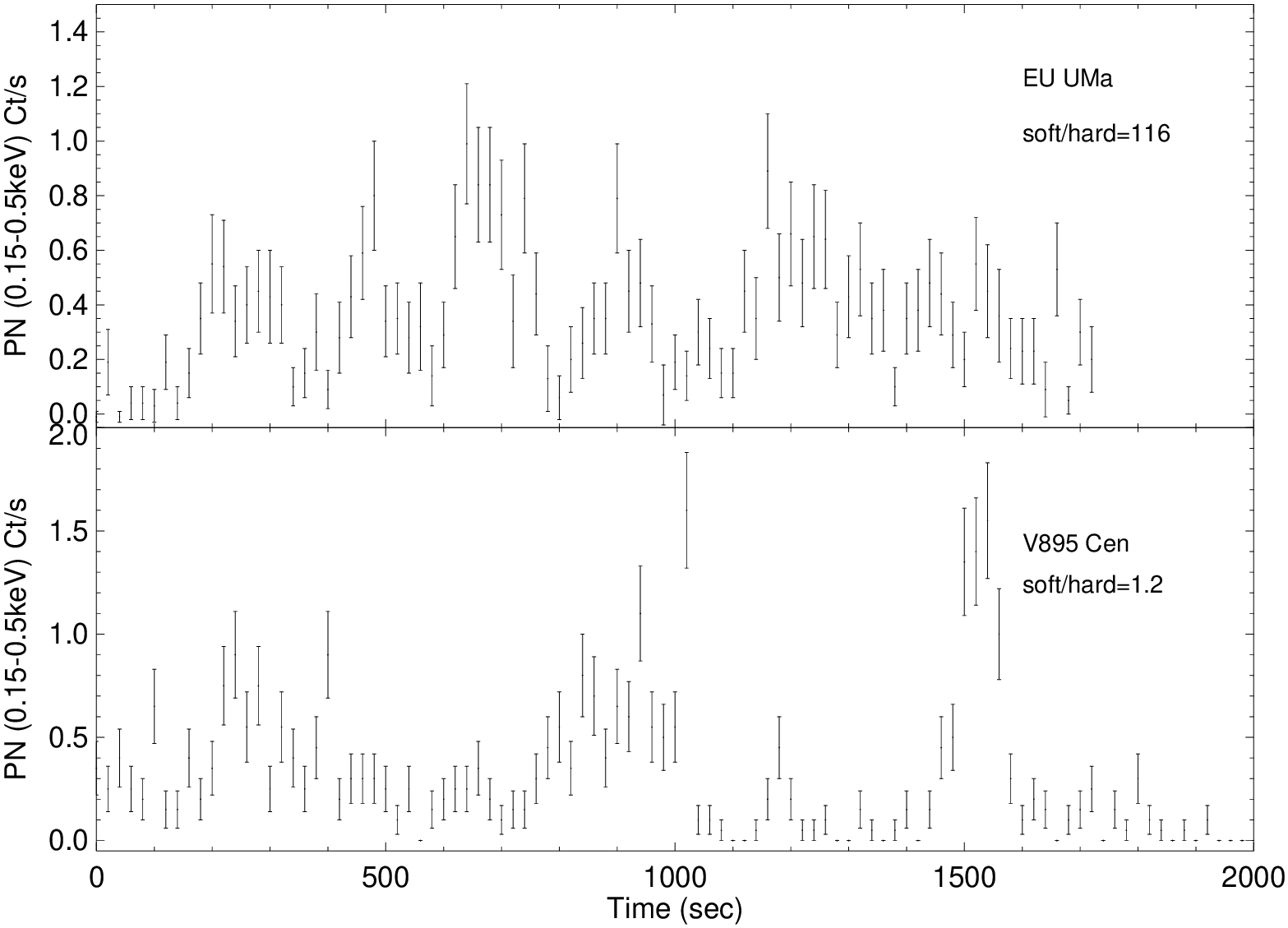}}
\end{picture}
\end{center}
\caption{The 0.15-0.5keV band light curves of EU UMa, which has a high
energy balance ratio, and V895 Cen, which has a low energy balance
ratio. Both systems show prominent flickering or flaring behaviour.}
\label{flickering}
\end{figure}

Flickering or flaring behaviour is seen in many polars. However, we
find that the degree of flaring does not necessarily affect the
resulting X-ray spectrum.  We show in Figure \ref{flickering} a
section of the 0.15--0.5keV light curves of EU UMa (energy balance
ratio 116 using {\xmm}) and V895 Cen (energy ratio 1.2): both systems
show flaring behaviour but very different ratios. Therefore, in
systems which show prominent flares or flickering behaviour, but low
energy balance ratios, the flares are likely to be due to enhancements
in the accretion flow, but which do not have sufficient density to
penetrate the photosphere of the white dwarf and therefore have buried
shocks.

\section{Systems with no soft X-ray component}
\label{nosoft}

Until now, one of the defining characteristics of polars has been the
presence of a strong soft X-ray component. It is therefore surprising
that we find 6 systems which showed at least one accretion pole which
did not show a distinct soft X-ray component at all. It is possible
that the reprocessed component is cool enough to have moved out of the
{\sl XMM-Newton} band as proposed by Heise \& Verbunt (1988).

\begin{figure}
\begin{center}
\setlength{\unitlength}{1cm}
\begin{picture}(8,11.5)
\put(-1.,-1.2){\includegraphics{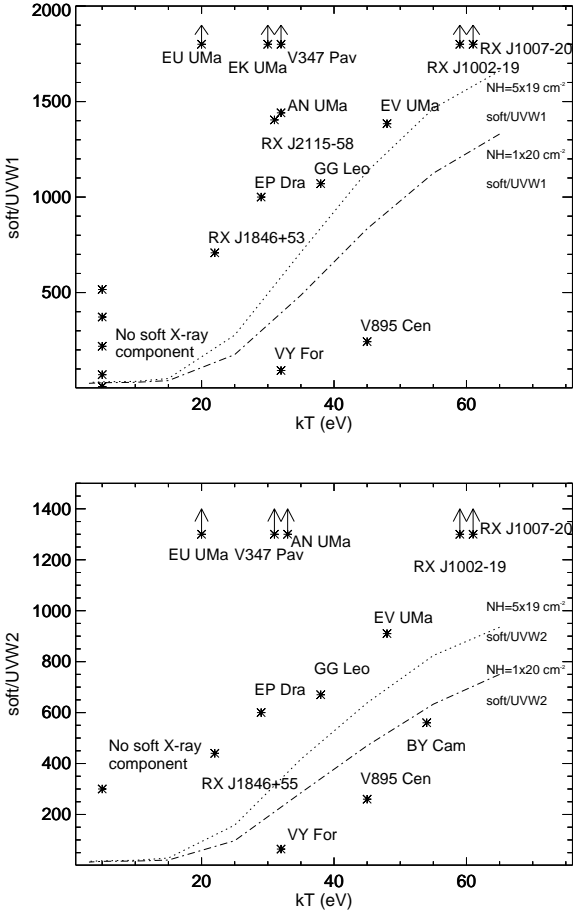}}
\end{picture}
\end{center}
\caption{We show the simulations of the expected flux ratios
0.15-0.5keV/UVW1 and 0.15-0.5keV/UVW2 assuming a blackbody of various
temperature and a shock component of 30keV. Their normalisations are
set so that the unabsorbed, bolometric luminosities give a ratio
consistent with the standard accretion model. We also include a
blackbody component of temperature 20000K to account for a white dwarf
of mass 0.7\Msun. We plot the best fit blackbody temperatures derived
from the X-ray spectra of each polar which has a soft X-ray component
and their 0.15-0.5keV/UV ratios. For those systems showing no soft
component we plot them, arbitrarily at $kT$=5eV.}
\label{ratio2}
\end{figure}

If this is the case then this may be reflected in the observed
0.15-0.5keV to UV flux ratios: for those systems which show no distinct
soft X-ray component this ratio should be lower than in those systems
which do. Firstly, we simulated the spectrum of a polar which had an
energy balance consistent with that predicted by the standard model,
with the temperature of the reprocessed component being in the range
3--65eV. We include in our model an unheated white dwarf of mass
0.7\Msun (and assume the Nauenberg 1972 mass-radius relationship) and
temperature 20000K (1.7eV) assuming a blackbody. We also include a
thermal bremsstrahlung component with temperature 30keV. We then
measured the expected flux in the 0.15--0.5keV band and in the UVW1
(an effective wavelength of 2910\AA) and UVW2 (2120\AA) OM filters. We
show the ratios for two values of absorption in Figure \ref{ratio2}
and tabulate the measured soft X-ray/UV ratios of our sample of {\xmm}
polars in Table \ref{softuv}.  Obviously this does not take into
account viewing angle affects or stream emission: for significant
stream emission these ratios would be greater.

To convert the observed count rate to flux we use the latest
conversion factor assuming a white dwarf spectrum: 1 ct/s in UVW1
implies a flux of 4.4$\times10^{-16}$ \ergscm \AA while 1 cts/s in
UVW2 gives 5.8$\times10^{-16}$ \ergscm \AA. There are some systems in
which the source was not detected in either UV filter (EU Cnc) or was
not detected in both (or only one UV filter was used). In the case of
BY Cam and RX J2115--58 (which showed one pole which did not show a
distinct soft component) we used the appropriate phase range.

For those systems with a distinct soft X-ray component, the
0.15-0.5keV/UV ratios are generally high, while in contrast, those
systems which do not show a distinct soft component, generally show
lower ratios.  Indeed, many of the soft X-ray systems show ratios
which are much higher than the results from the simulation. If the
temperature of the white dwarf in these systems is hotter than 20000K
the ratios will be higher: for a blackbody of temperature 40000K, the
0.15--0.5keV/UV ratios increase by a factor of 3.5, while for a
temperature of 60000K they increases by a factor of 56. Further, if
the system shows a high soft X-ray excess, or has a significant UV
stream component then this will be reflected in high 0.15-0.5keV/UV
ratios.

Of the `soft' X-ray systems only VY For shows a relatively low
ratio. It is not clear why this is the case -- fitting its X-ray
spectrum does not show a particularly high absorption. It is
interesting to note, however, that Beuermann et al (1989) and Cropper
(1997) suggested that its main accretion pole is hidden from view and
the visible pole was the secondary pole. In a sample such as ours we
may expect more than one system to have a `hidden' pole.

Care should be taken when interpreting these results since, as we
pointed out above, we assume stream emission does not contribute to
the UV flux and we do not take into account optical thickness
affects. However, we find that overall, those systems which do not
show a distinct soft X-ray component give lower (0.15-0.5keV)/UV flux
ratios, consistent with our view that the reprocessed component has
been shifted from soft X-ray energies to UV energies.

\begin{table}
\begin{center}
\begin{tabular}{lrr}
\hline
Source & 0.15--0.5keV/ & 0.15--0.5keV/ \\
       & UVW1 & UVW2 \\
\hline
CE Gru & 520 & 300 \\
V349 Pav & 220 & \\
V1500 Cyg & 10 & \\
BY Cam & 70 & \\
RX J2115--58 & 370 & \\
\hline
BY Cam & & 810 \\
RX J1007--2016 & 7800 & 4400 \\
EV UMa & 1400 & 910 \\
RX J1002--1925 & 2600 & 1800 \\
V895 Cen & 240 & 260 \\
V347 Pav & 3600 & 1800 \\
RX J2115--58 & 1400 & \\
AN UMa & 1400 & 1700 \\
EK UMa & 15000 & \\
GG Leo & 1100 & 670 \\
EU UMa & 610 & 3000\\
EP Dra & 1000 & 600 \\
VY For & 90 & 64 \\
RX J1846+53 & 710 & 440\\
\hline
\end{tabular}
\end{center}
\caption{The observed flux ratios 0.15--0.5keV/UVW1 and
0.15--0.5keV/UVW2 for systems which did not show a distinct soft X-ray
component (top) and those which did (bottom).}
\label{softuv}
\end{table}

In the scenario of Heise \& Verbunt (1988), those systems which show a
soft X-ray component are accreting (at least some) dense blobs of
material. However, for most polars we find that the energy balance is
consistent with the standard accretion model -- which does contain a
significant contribution from dense blobs. Therefore, Heise \& Verbunt
(1988) are not correct in saying that a soft X-ray component {\sl
requires} blobs. 

The factors that may affect the temperature of the soft component are
$M_{wd}$ (since that sets the maximum temperature in the shock),
$\dot{M}$ (since this sets the height of the shock) and the magnetic
field strength (since this sets amount of cooling due to cyclotron
radiation).

\section{Conclusions}

Our survey of polars using {\xmm} has shown that most systems have
X-ray spectra which give relatively low soft-to-hard X-ray
ratios. This is in contrast to the results of Ramsay et al (1994) who
used Rev0 {\ros} data and found that the energy balance ratio was
biased towards high ratios. We have re-examined all the polars
observed in the pointed mode using the {\ros} PSPC. Using Rev2
calibrated data we find that the energy ratio has a similar
distribution to that of the {\xmm} sample, with only a slight increase
in the relative number of systems having a high ratio. We conclude
that the results of Ramsay et al (1994) which showed many systems had
high ratios were due to their data being less well calibrated than is
now possible.

\begin{figure}
\begin{center}
\setlength{\unitlength}{1cm}
\begin{picture}(8,8.5)
\put(-0.5,-0.3){\includegraphics{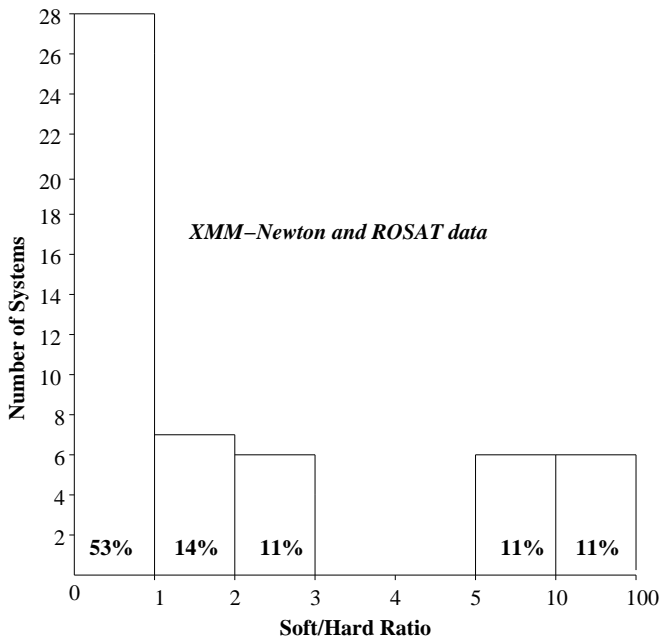}}
\end{picture}
\end{center}
\caption{The soft/hard ratio for the source included in our {\xmm}
survey and all the polars which were observed in the {\ros} PSPC
pointed programme.}
\label{ratio.rosat.xmm}
\end{figure}

We show the distribution of the soft-to-hard ratio using the combined
{\xmm} and {\ros} samples in Figure \ref{ratio.rosat.xmm}. Most
systems show low ratios: once consideration of affects such as
reflection of hard X-rays from the surface of the white dwarf, the
correction for optical thickness effects and the cyclotron component
are taken into account, most systems are likely to be consistent with
the standard accretion model. However, there are still a number of
systems which show large excesses. We have explored the physical
reasons for such an excess and speculate that the orientation between
the magnetic field axis and the stream as it meets the magnetic field
of the white dwarf may be a likely parameter.

We find that systems show ratios which are related to their intensity:
when an individual systems is bright it shows a higher ratio. However,
we find no evidence that their actual luminosity sets the ratio, so
that more luminous systems can have the same ratio as less luminous
systems. 

Another surprising finding from our survey is the number of systems
which do not show any distinct soft X-ray component. We suggest that
this is due to the reprocessed component being shifted to lower
energies and hence out of the {\xmm} X-ray window. Further study of
what determines the temperature of the reprocessed component is
strongly encouraged.

\section{acknowledgments}

This paper is based on observations obtained using {\sl XMM-Newton},
an ESA science mission with instruments and contributions directly
funded by ESA Member States and the USA (NASA). These observations
were part of the OM guaranteed time programme. We have also made use
of the {\ros} data archive at MPE, Germany. We thank Kinwah Wu for
useful discussions.


\begin{thebibliography}{99}

\bibitem{}Arnaud K. A., 1996, Astronomical Data Analysis Software and 
Systems V, eds. Jacoby G. and Barnes J., p17, ASP Conf. Series volume 101.
\bibitem{}Beuermann K., Burwitz V., 1995, ASP Conf Series, Vol 85, 99
\bibitem{}Beuermann K., Thomas H.-C., Giommi P., Tagliaferri G., 
Schwope A. D., 1989, A\&A, 219, L7
\bibitem{}Bridge C., Cropper M., Ramsay G., Perryman M. A. C., de
Bruijne J. H. J., Favata F., Peacock A., Rando N., Reynolds
A. P., 2002, MNRAS, 336, 1229
\bibitem{}Bridge C. M., Cropper M., Ramsay G., de Bruijne J. H.,
Reynolds A. P., Perryman M. A. C., 2003, MNRAS, 341, 863
\bibitem{}Cropper M., 1988, MNRAS, 231, 597
\bibitem{}Cropper M., 1990, Space Sci Rev, 54, 195
\bibitem{}Cropper M., 1997, MNRAS, 289, 21
\bibitem{}Cropper M., Wickramasinghe D. T., 1993, MNRAS, 260, 696
\bibitem{}Cropper M., Wu K., Ramsay G., Kocabiyik A., 1999, 
MNRAS, 306, 684
\bibitem{}Cropper M., Wu K., Ramsay G., 2000, New Astron. Rev., 44, 57 
\bibitem{}Cropper M., Ramsay G., Marsh T., 2003, In Proc Cape Town Workshop
on mCVs, ASP Conf Series, astro-ph/0303265
\bibitem{}Done C., Magdziarz P., 1998, MNRAS, 298, 737
\bibitem{}Fabian A., Pringle J., Rees M., 1976, MNRAS, 175, 32
\bibitem{}Frank J., King A. R., Lasota J. -P., 1988, A\&A, 193, 113
\bibitem{}Frank J., King A., Raine D., 2002, Accretion Power in
Astrophysics, 3rd Edition, Cambridge University Press
\bibitem{}Greeley B. W., Blair W. P., Long K. S., Raymond J. C.,
1999, ApJ, 1999, 513, 491
\bibitem{}Heise J., 1995, In Cape Workshop on Magnetic cataclysmic
variables, ASP Conf Ser, 85, Ed. D. A. H. Buckley \& B. Warner, 162
\bibitem{}Heise J., Verbunt F., 1988, A\&A, 189, 112
\bibitem{}Jansen F., et al, 2001, A\&A, 365, L1
\bibitem{}King A. R., 2000, ApJ, 541, 306
\bibitem{}King A. R., Lasota J. P., 1979, MNRAS, 188, 653
\bibitem{}King A. R., Lasota J. P., 1980, MNRAS, 191, 721
\bibitem{}Kuijpers J., Pringle J. E., 1982, A\&A, 114, L4
\bibitem{}Lamb D. Q., Masters A. R., 1979, ApJ, 234, 117
\bibitem{}Litchfield S. J., King A. R., 1990, MNRAS, 247, 200
\bibitem{}Mason K. O., et al 2001, A\&A, 365, L36
\bibitem{}Nauenberg M., 1972, ApJ, 175, 417
\bibitem{}Pandel D., Cordova F., Shirey R., Ramsay G., Cropper
M., Mason K., Much R., Kilkenny D., 2002, MNRAS, 332, 116
\bibitem{}Papaloizou J. C. B., Pringle J. E., MacDonald J., 1982,
MNRAS, 198, 215
\bibitem{}Ramsay G., Mason K. O., Cropper M., Watson M. G.,
Clayton K. L., 1994, MNRAS, 270, 692
\bibitem{}Ramsay G., Cropper M., Mason K. O., 1995, MNRAS, 276, 1382
\bibitem{}Ramsay G., Cropper M., Cordova F., Mason K., Much R.,
Pandel D., Shirey R., 2001, MNRAS, 326, L27
\bibitem{}Ramsay G., Cropper M., 2002a, MNRAS, 334, 805
\bibitem{}Ramsay G., Cropper M., 2002b, MNRAS, 335, 918
\bibitem{}Ramsay G., Cropper M., 2003a, MNRAS, 338, 219
\bibitem{}Ramsay G., Cropper M., 2003b, In Proc Cape Town Workshop
on mCVs, ASP Conf Series, astro-ph/0301609
\bibitem{}Ramsay G., Cropper M., Mason K., Cordova F., Priedhorsky
W., accepted, MNRAS
\bibitem{}Raymond J. C., Davis R. J., Hartmann L., Matilsky T. A., 
Black J. H., Dupree A. K., Gursky H., 1979, ApJ, 230, L95
\bibitem{}Reinsch K., Burwitz V., Beuermann K., Thomas H. -C.,
1999, ASP Conf Series, Vol 157, 187
\bibitem{}Schwope A., Schwarz R., Sirk M., Howell S. B., 2001,
A\&A, 2001, 375, 419
\bibitem{}Sokoloski J. L., Bildstein L., Ho W. C. G., 2001, MNRAS,
326, 553
\bibitem{}Str\"{u}der L., et al, 2001, 365, L18
\bibitem{}Thompson A. M., Cawthorne T. V., 1987, MNRAS, 224, 425
\bibitem{}Turner M., et al 2001, A\&A, 365, L27
\bibitem{}Warner B., 1995, Cataclysmic variable stars, Cambridge
Univ. Press, Cambridge
\bibitem{}Wickramasinghe D. T., Ferrario L., 2000, PASP, 112, 873
\bibitem{}Williams G., King A. R., Brooker J. R. E., 1987, MNRAS,
226, 725
\bibitem{}Wu K., 2000, Space Science Reviews, 93, 611
\end{thebibliography}
\end{document}